\documentclass[conference]{IEEEtran}
\usepackage{multirow}
\usepackage{graphicx}
\usepackage[export]{adjustbox}
\usepackage{setspace}
\usepackage{array}
\newcommand{\RN}[1]{\textup{\uppercase\expandafter{\romannumeral#1}}}
\begin{document}

\title{Towards Successful Social Media Advertising: Predicting the Influence of Commercial Tweets}

\author
{
\IEEEauthorblockN{Renhao Cui}
\IEEEauthorblockA{
The Ohio State University\\
Columbus, OH 43210\\
cui.182@osu.edu
}
\and
\IEEEauthorblockN{Gagan Agrawal}
\IEEEauthorblockA{
The Ohio State University\\
Columbus, OH 43210\\
agrawal.28@osu.edu
}
\and
\IEEEauthorblockN{Rajiv Ramnath}
\IEEEauthorblockA{
The Ohio State University\\
Columbus, OH 43210\\
ramnath.6@osu.edu
}
}
\maketitle

\begin{abstract}
Businesses communicate using Twitter for a variety of reasons -- to raise awareness of their brands,
to market new products, to respond to community comments, and to connect with their customers and potential customers in a targeted manner.
For businesses to do this effectively, they need to understand which content and structural elements about a tweet make it influential, that is, widely liked, followed,  and retweeted.
This paper presents a systematic methodology for analyzing commercial tweets, and predicting the influence on their readers.
Our model, which use a combination of decoration and meta features, outperforms the prediction
ability of the baseline model as well as the tweet embedding model.
Further, in order to demonstrate a practical use of this work, we show how an unsuccessful tweet may be engineered (for example, reworded) to increase its potential for success.
\end{abstract}

\section{Introduction}
The rapid growth of social media is driving the increased use of social platforms for advertising.
Many companies have official accounts on social media platforms to maintain customer relationships, spread news, and attract more attention.
In fact, companies have used their official accounts on Twitter to post commercial tweets that are primarily visible to their followers.
For example, {\em ``Is your New Year's resolution to travel more? Check out these up-and-coming destinations - https://t.co/I36OS2h
BnF https://t.co/rju66wtv5a''} is an advertisement tweet posted by the {\em Travel Channel}.

Analysis of Twitter posts has attracted significant attention from the data-mining community in recent years.
The massive volume, real-time nature, large geographical coverage, and public availability of Twitter data have led to this heightened interest.
Mining Twitter data has been demonstrated to be useful for tasks such as earthquake detection~\cite{sakaki2010earthquake}, stock market
prediction~\cite{bollen2011twitter}, public health applications \cite{paul2011you}, and open-domain event extraction~\cite{ritter2012open}.

In using social media to advertise their products and promotions, companies are seeking more engagement from their readers as part of maintaining an effective online strategy.
This need has led to a new class of services being offered to help the companies build stronger relationships with their customers.
Social Customer Relationship Management (CRM), compared with traditional CRM,
aims to provide a closer and more direct communication between the company and its customers through different social platforms.
More broadly, increasing the effectiveness of advertising through social media continues to be an intriguing and open question for various corporations.
Many approaches exist to measure the influence of an individual account on a social platform,
such as the widely used Klout Score \cite{rao2015klout}.
For a single company, the focus is more often on the effectiveness of its messages propagation.
Thus, there is now a need to understand how to raise the influence of a particular post.

This chapter shows that the influence of a particular commercial post can be measured, predicted, and made more effective.
The techniques presented here may be used within a system to help companies craft commercial messages for social platforms,
with the goal of maximizing the influence of the posts on specific audiences.
In order to improve the writing of a commercial post, predicting the potential influence and effectiveness of a given text is an essential and important first step.

The primary contribution is to answer the following question:
{\em ``Can we learn what makes an effective advertising post on Twitter?''}
In doing so, we address the following challenges:
\begin{itemize}
\item How can we quantify whether a given commercial post on Twitter has been successful?
\item How can we distinguish the influencing (decorative) elements of a commercial post from its inherent meaning?
\item What features best model the composition of a tweet with respect to its influence?
\item What is the specific effect of each of the specific features in improving the influence of the tweet?
\item Can a tweet be engineered to have more influence?
\end{itemize}

We consider a commercial post to be successful if it can generate enough {\em influence}.
To measure influence, we use the direct reactions that a tweet gets from its readers -- such as retweeting and marking as favorite.
We focus on commercial posts from the official accounts of various companies (brands).
Although  pictures and/or videos can be included to enrich a post, we believe that the
text content provides the most important and straightforward information.
Note that the product or promotion information is usually determined before crafting the post, so we shift the focus to the other influencing elements of the post.
First, we label the commercial tweet based on the influence it generated.
Then, we extract a small set of features to capture the structure and comprehensive representation of a commercial post.
More specifically, the feature set is designed to show the construction of a post and it does not include the core information that is related to the promotion or product.
Next, we address the problem of predicting whether a commercial post would be successful through a binary classification model given the feature set.
In addition, we conduct a feature analysis to determine which features have the most impact on the influence prediction.
We provide a case study that shows the potential usage of the prediction system.

To the best of our knowledge, this is the first model that seeks to analyze and predict the performance of commercial social media posts.
We believe that this work will serve as an essential foundation for advertising-related social media analysis.
\section{Related Work}

Advertising through social media is growing rapidly and is drawing more attention.
Yin et al. \cite{yin2012discovering} used the concept of a propagation tree to reveal patterns of advertisement propagation and present a set of metrics to measure the effectiveness of an advertisement in terms of extent of its propagation.
In contrast, our focus is on tweet content and measuring the advertisement by its influence on the readers.
Li et al. \cite{li2012diffusion} proposed a diffusion mechanism for advertisement delivery through microblog platforms, based on a set of user-related features.
Our goal is to model the influence of a commercial tweet based on static textual features before it has been posted.

In the last decade, researchers have examined the influence of specific users and their posts through social media and attempted to understand how to quantify such influence.
Anger et al. \cite{anger2011measuring} looked into the indicators of influence on Twitter for an individual user.
Bakshy et al. \cite{bakshy2011everyone} conducted a study quantifying the influence of Twitter users by tracking the diffusion of their posts through reposts.
Cha et al. \cite{cha2010measuring} also focused on user influence and proposed a link-based model to measure influence.
Ye and Wu \cite{ye2010measuring} proposed a model to measure message propagation and its social influence through Twitter,
as well as the longitudinal influence over time and across users.
Unlike these network-based models or diffusion models that track the spread of tweets,
we construct a simple matrix that checks only direct reactions to tweets by their readers (such as favorites and retweets).
Moreover, in order to improve a given post, we focus on the specific tweet, instead of the identity of the author.

Popularity prediction also has attracted much interest and among many models, predicting retweets has been the most common one.
Most efforts (such as \cite{petrovic2011rt} and \cite{naveed2011bad}) utilized simple surface features of tweets to predict retweeting.
Peng et al. \cite{pengretweet} also included relationship features,
and Yang et al. \cite{yang2010understanding} added trace and temporal features to build a factor graph model.
Zaman et al. \cite{zaman2010predicting} and Gao et al. \cite{gao2015modeling} predicted future popularity by observing the dynamics of retweeting,
while Xu et al. \cite{xu2012analyzing} and Lee et al. \cite{lee2014will} focused on retweet activities on certain users.
In addition to retweet prediction, Artzi et al. \cite{artzi2012predicting} predicted whether a tweet will receive replies from its readers,
and Suh et al. \cite{suh2010want} showed the relation between certain features and the retweet rate.
In contrast to these efforts, which focused on a single reaction (such as retweeting) as the measurement, our work creates a comprehensive metric to measure the popularity (influence) of commercial tweets.
Furthermore, our model captures only the structural and decoration elements of the post, rather than checking every detailed word of it.

Distributed representation has become popular in text-related research.
Mikolov et al. \cite{mikolov2013efficient} initiated the work on representing words with lower dimension vectors, which are trained to predict context words given the current word.
Le et al. \cite{le2014distributed} leveraged \cite{mikolov2013efficient} to represent a paragraph using a dense vector,
which is trained to predict words in the paragraph given the paragraph itself.
The idea of dense representation also has been brought to tweet-related tasks.
Tang et al. \cite{tang2014learning} built a word embedding for Twitter sentiment classification.
Given the informal use of words in tweets, two character-based {\em tweet2vec} models have been introduced: 
Dhingra et al. \cite{dhingra2016tweet2vec} constructed a tweet vector representation to predict hashtags,
and Vosoughi et al. \cite{vosoughi2016tweet2vec} use a CNN-LSTM encoder-decoder model to generate tweet embeddings.
\section{Influence Prediction}
This section describes the process of labeling the influence of commercial tweets, extracting features and building classification models.

\subsection{Data labeling}
In order to  classify commercial tweets as {\em successful}  or {\em unsuccessful}, we first need to quantify the {\em influence} of a tweet.
The influence of a commercial tweet can be represented by the level of engagement from the readers.
Retweets and  marking tweets as favorites are the most widely used functions that allow a reader to express his or her interest or excitement
about the tweet.
Thus, counts of retweets and favorites can be used as direct measurements of reader engagement and they have been used in some models as the indicator for tweet influence \cite{cha2010measuring,ye2010measuring}.

\subsubsection{Influence Score}
Our work combines the count of retweets and the count of favorites in order to measure tweet influence.
Both reactions reflect the interest of the user after reading the tweet.
We want the counts of each reaction to have equal impact in determining the influence.
However, users retweet and mark as favorite with different frequencies, which leads to different scales for the two counts.
To balance the scale of the two counts, we compute the ratio of favorite-to-retweet counts across all tweets in the dataset, as shown in Figure 1.

\begin{figure*}[htb]
\centering
\includegraphics[scale=0.5]{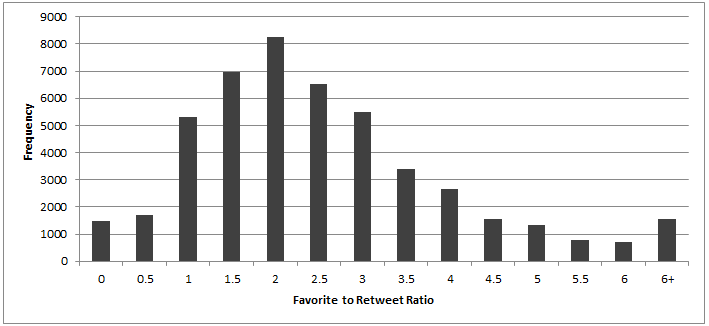}
\caption{Favorite-to-retweet ratio}
\end{figure*}

Most tweets have favorite-to-retweet ratios around $2$, and the mean of the ratios across the dataset is $2.5$.
Thus, we empirically weight retweet count by $2$ to ensure that retweets and favorites have equal influence in the final measurement.
In fact, we think this multiplier may reflect the fact that marking a favorite requires one click, while retweeting requires two clicks.

However, retweet and favorite counts are highly influenced by the popularity of the author account, which can be identified as the number of followers.
Note that we are not interested in the absolute influence created by a post, but the relative influence that the author account is able to generate.
To create a normalized influence for all general commercial posts,
we eliminate the impact of account popularity by normalizing the score by the number of followers of the account.
Therefore, the influence score in our work is calculated as:
\begin{eqnarray}
\frac{2\times RetweetCount + FavoriteCount}{FollowerCount} \nonumber
\end{eqnarray}
Because the influence score is normalized by the number of followers, we include only direct reactions of a tweet from its readers for influence measurement, 
rather than tracking the propagation of the tweet.

Retweet, favorite, and follower counts are all dynamic attributes in the sense that, for a given tweet or account, they change with time.
Further, tweets are time-sensitive, and the attention they receive only lasts for a short period of time.
Willis et al. \cite{willis2015mapping} have shown that most retweets happen in the first 20 hours after the original post.
Thus, to provide stable data, we record account information at the time of tweet posting and collect tweet data three weeks after posting.

\subsubsection{Separation from Inherent Meanings}
A basic property of commercial posts is that they are used to spread certain information.
In general, companies have decided the content of the promotion or products in advance of constructing the tweet, and such information should be considered as fixed. 
Therefore, the inherent meaning of the post should not be included in order to predict the influence of spreading such information.
Thus, we design a process to distinguish the core meaning of the post from other decoration features. 
Although both elements could affect the successfulness of the commercial post, we want to focus the study on only the decoration features.
This is done in the following manner.

First, we conduct a part-of-speech (POS) tagging \cite{gimpel2011part} on the tweet, and extract nouns, verbs, adverbs, and adjectives.
These words or phrases are considered as key words. 
In most cases, these key words are the carrier of the inherent meaning of the post.
Next, we group the tweets given these key words, using certain clustering methods.
The goal is to group posts that are writing about similar products or promotions.
Given that the core meaning of the posts is similar in some ways, the model can study the relation between the remaining decoration features and the generated influence.

We also note that the overall distribution of the influence scores is  biased toward smaller scores.
Thus, if we use the influence score directly and define the task as a regression problem,
the result may also be skewed toward very small scores.
To remove this bias, we treat the labeling process as a binary classification problem,  where the top $50\%$ of the tweets (higher scores) are labeled as positive, and the bottom 50\% are labeled as negative.
Because the labeling process is performed independently for each group generated from the previous step, 
it is able to reduce the impact of the influence caused by the inherent meaning of the post from the labels.

\subsection{Classification model}

\begin{table*}[httb]
\centering
\caption{Decoration features}
\begin{tabular}{lc|lc}
\hline
\textbf{Feature}&\textbf{Values}&\textbf{Feature}&\textbf{Values}\\
\hline
\hline
\textbf{Complexity Features}&&\textbf{Post Meta Features}&\\
Tweet length & Integer & Day of week& Categorical\\
Readability score & Continuous & Time of Day& Categorical\\
Parse tree depth & Continuous &\textbf{Mention Features}&\\
Parse tree head count & Continuous & Verified username & Binary\\
\textbf{Element Features}&&Username follower count & Continuous\\
Usernames & Binary &\textbf{Punctuation Features}&\\
URLs & Binary &Question marks & Binary\\
Hashtags & Binary &Exclamation marks&Binary\\
\textbf{Author Meta Features} & &\textbf{Other Features}&\\
Post count & Continuous &Contain digits&Binary\\
Favorite count & Continuous &POS dist (3)& [0, 1]\\
Listed count & Continuous &Sentiment& Continuous\\
\hline
\end{tabular}
\end{table*}

As mentioned above, and given the binary labeling of tweets, we use a binary classifier for the prediction model.
To explore differences in impact of the proposed features from general contextual features, we use the typical {\em n-gram} model as the baseline.
More specifically, the baseline comprises {\em n-gram} features up to length $5$, including all tokens that appear in the dataset more than once.
This approach is the de-facto standard for text classification tasks such as sentiment classification and topic categorization~\cite{agichtein2008finding}.
Unlike previous work \cite{pang2002thumbs,agichtein2008finding}, we find that Maximum Entropy (MaxEnt) \cite{manning1999foundations} works better 
than Support Vector Machine (SVM) \cite{joachims1998text} with the baseline model for classifying the dataset.
We also apply the state-of-the-art character-based tweet embedding model \cite{dhingra2016tweet2vec} as a comparison.
The tweet embedding model is trained to predict the hashtags.
Hashtags in commercial tweets generally contain information about products or promotions, which makes the embedding model a good fit for commercial posts.
Because SVM performs better generally, we apply it to the tweet embedding model as well as to the proposed comprehensive set of features that we will describe in the following sections.

\subsection{Features}
Although our ultimate goal is to improve a tweet to be more influential, we first focus on  predicting the influence of a commercial tweet.
Unlike predicting the cascade of retweeting \cite{cheng2014can}, we do not include any observation of the diffusion of the post, but extract features that are available instantly.
More specifically, given the broad message that is being captured by the tweet,
we want to model how the high-level structural and meta information impact its influence.

We propose a set of features that works as a high-level representation of the post.
Because the features do not include the inherent meaning of the post, we name them {\em decoration features}.
The proposed features are built to capture the structural and syntactic characteristics of a post, and also includes certain pertinent information about the posting account.
Nasir et al. \cite{naveed2011bad} have looked  into similar features in order to predict retweeting for general posts.
Building on their work, we construct the feature set shown in Table \RN{1} for commercial post influence prediction.

\subsubsection{Element Features}
Usernames, hashtags, and links (URLs) are often used to deliver important information, especially considering the length limitation of tweets.

{\em  Usernames} mentioned in the tweet are usually used to refer to specific users, or alternatively, used
 to send the tweet to the users.
It is a common way to attract readers by mentioning a popular user in the post.

{\em Hashtags} serve to identify a certain topic and are often treated as symbols across tweets that share the same idea.
For commercial tweets, it is also common to use hashtags as representations of certain products or events.
Thus, the information carried by the hashtag is critical and has a big impact on the influence of the tweet.

{\em URLs} work as an extension to the tweet in order to include detailed and richer information.
For commercial posts, they play a critical role in  pointing readers to additional information and details.
Therefore, they can potentially increase the chance of being retweeted or marked as a favorite for the tweet.

Note that the intention of the system is not to alter the inherent meaning of a post, 
and to look for general features that affect the influence of commercial tweets, 
so we do not look into the actual content or the semantics of these elements.
Instead, these features are represented as binary indicators, and these elements are tokenized for other processes.

\subsubsection{Punctuation Features}
Rhetorical questions are popular hooks for commercial posts, and {\em question marks} serve to demarcate such hooks.
{\em Exclamation marks} are often used to express strong emotions.
Commercial tweets are written more formally than general tweets; thus, the use of such punctuations marks is deliberate.
We use a binary feature for each punctuation mark to represent its existence in a tweet.

\subsubsection{Complexity Features}
The complexity of a tweet indicates the ease (or difficulty) of  reading, understanding,  and interpreting
 the content.
We measure such complexity using four features.

{\em Tweet length} is a straightforward indicator of complexity.
The $140$-character limitation is applied to the number of characters of a single tweet,
but the tense of a verb, the proper name, and even the URL, skew the count of characters.
Therefore, instead of counting characters, we use the number of tokens to represent the length of a tweet.
It is used both as a feature and the normalization factor of other features.
The analysis of our commercial tweet dataset showed that the average number of tokens is $15.2$, with a standard deviation of $5.1$.
This shows a significant level of variation in tweet length, and thus it has the potential to become an indicator.

{\em Readability} is a measure of the difficulty of reading and comprehending a piece of text.
It has been used as an indicator for the quality of social media content \cite{agichtein2008finding}.
In this work, we use the {\em Coleman-Liau Index} \cite{coleman1975computer} as the readability feature.
The score is calculated as:
\begin{eqnarray}
CLI = 0.0588L - 0.296S -15.8 \nonumber
\end{eqnarray}
where $L$ is the average number of letters per 100 words, and $S$ is the average number of sentences per 100 words.
The resulting score is an approximation of the U.S. grade level needed to understand the text.
Similar to tweet length, readability score captures the surface complexity of the post.

The dependency parse tree of a tweet shows the structure of the text, with both word- and phrase-level relations.
We build the dependency parse tree for each tweet using the Twitter-specific model proposed by Kong et al.~\cite{kong2014dependency}.
The parse tree is able to capture the intrinsic and structural property of the tweet.
Given such a parse tree, the depth of a tweet is the number of levels starting from the root node to the bottom of the tweet parse tree.
Thus, {\em parse tree depth} can be used as a feature to measure the dependency complexity of the tweet.
{\em Parse tree head count} is the number of syntactic roots contained in the tweet parse tree.
Each root leads to an individual fragment of the tweet, which is considered to be a complete and meaningful portion.
It is not necessarily equal to the number of sentences because complete fragments can be separated in many ways.
In general, a commercial tweet contains a single topic.
Therefore, a tweet with more heads in the parse tree tends to have a higher density of information about the topic.
We use the head count to serve as a feature to measure the density complexity of the tweet.
Note that both parse tree depth and head count are normalized by the length of the tweet.

\subsubsection{Mentions Features}
As described above, the usernames mentioned in commercial tweets could help  attract readers.
In most cases, influence is driven by the popularity of the usernames and their linked accounts.
Therefore, in addition to the existence of usernames, we also use the popularity of the usernames' accounts as a feature.
We use two attributes of the mentioned username to measure its popularity: whether it is a verified account, and its follower count.
{\em Verified usernames} belong to persons whose accounts have been certified as {\em genuine}, which often are associated with ``famous people'' \cite{marwick2011see}.
The verification of an account indicates its popularity, and the username verification feature is set to have a binary value. 
However, only a small portion of Twitter accounts are verified and thus the applicability of this indicator is limited.
{\em Username follower count}, on the other hand, is a quantifiable estimator of the popularity available for all accounts.
The username follower count feature is calculated as the average number of followers across all usernames mentioned in the post.

\subsubsection{Meta Features of the Post}
Previous work has shown  that the posting time of a tweet influences the retweet or response
potential~\cite{artzi2012predicting}.
Therefore, we include both the {\em day of week} and the {\em time of day} as meta features for the tweet.
For both features, we use the local time, and further map the time into four periods (consisting of six hours each).

\subsubsection{Meta Features of the Author}
The author of a tweet has been shown to have a significant impact on the influence of general tweets \cite{bakshy2011everyone,cha2010measuring}.
We want to extend such impact to the relation between official accounts and commercial tweets.
To prevent over-fitting and to make the model more general, we chose attributes that do not reveal the identity of the author account.
{\em Post count} is the number of tweets that this official account has posted,
while {\em favorite count} is the number of tweets that this account has marked as favorites.
Both counts represent the vitality of the account.
{\em Listed count} is the number of users that include this account in their interest lists and it can indicate the popularity of this official account on the social platform.
To eliminate the impact associated with the history of the account on these attributes,
we normalize the post count by the number of days between the registration of the account and the posting date, 
normalize favorite count by the post count, and normalize listed count by the number of followers of this account.

\subsubsection{Other Features}
Sentiment classification has been studied comprehensively \cite{pang2002thumbs} and has been used in previous tweet
influence analysis efforts \cite{naveed2011bad}.
The {\em sentiment} of a tweet is a potential factor that induces the attention of the readers.
Because commercial tweets mostly contain products or event-related information, they usually convey a nonnegative sentiment in their text.
Thus, sentiment may provide less differentiation ability, but the numeric value of the score can still be used as a measurement of the strength of the corresponding sentiment.
We use the Affective Norms for English Words (ANEW),
a microblog-based sentiment word list \cite{nielsen2011new}, to generate the sentiment score.
Because the output sentiment score is a summation of all the scores assigned to each word, we normalize the output score by the length of the tweet.

A POS tagger labels each word with a certain usage type, given the context of the word.
The POS tag feature has been shown to be useful for many types of social text mining tasks \cite{pang2002thumbs,kouloumpis2011twitter}.
To capture the critical information of a commercial post, we use $5$ from the list of $25$ Twitter-specific POS tags \cite{gimpel2011part}: common noun,
proper noun, verb, adjective, and adverb.
These five tags are then clustered into three POS categories: 1.) common noun and proper noun as the noun category; 2.) adjective and adverb as the descriptor
category; and 3.) verb as the verb category.
Similar to extracting meaningful content for labeling, we use the Gimpel \& Owoputi Twitter POS tagger to generate the sequence of POS tags for each tweet~\cite{gimpel2011part}.
To represent the writing style of the post, {\em POS distribution} features are then calculated as the normalized POS category counts across all three categories.

Digits in the commercial post often carry meaningful information -- such as useful statistics or emphasis on key ideas.
The binary feature of {\em containing digits} captures this role of digits in commercial tweets.
\section{Experiments}

\begin{table*}[htp]
\centering
\caption{Brands collected for the dataset}
\begin{tabular}{ccccc}
\hline
Gap &Amazon & Gilt & BlackBerry & Google\\
Nordstrom &Best Buy &Jeep & KraftFoods & Disney\\
AT\&T & Applebee's & Dell & Comcast & LEVIS\\
Macy's & AppStore (Apple) & JC Penney & Delta & H\&M\\
Starbucks & Travel Channel & FedEx & Yahoo&Motorola\\
SamsungMobile & Microsoft & Target & Sears&AmericanExpress\\
Netflix & GEICO & WholeFoods &&\\
\hline
\end{tabular}
\end{table*}

In this section, we describe our experiments that show the performance of different models.
All evaluations are performed using five-fold cross validation tests.
\begin{figure*}[htb]
\centering
\includegraphics[scale=0.38]{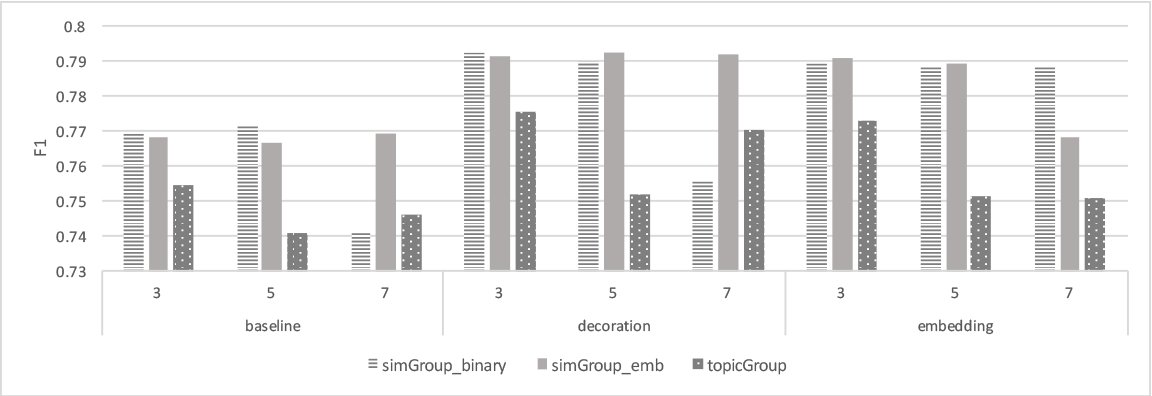}
\caption{Model performance given different labeling groups}
\end{figure*}
\subsection{Data preparation}
We build a commercial tweet dataset that contains originating tweets (i.e., no replies or retweets) posted by the official accounts of $33$ companies (Table \RN{2}).
During a 12-month period, 63,421 tweets were collected using the public Twitter API.
We found that most official accounts are very active in communicating with customers through retweeting and replies, 
but they are cautious in posting original commercial tweets, which generates a limited amount of useful data for the experiment.
The source code and dataset for the experiment have been made available.\footnote{https://goo.gl/Y1LFLA}

Outliers are removed in two steps.
Certain announcements, for example, the release of a new iPhone,
have an outsized influence simply because of the information, and thus the representation may not be the reason for their success.
Tweets that are related to such major announcements or events are found and excluded by keywords.
Other attributes may also cause an unpredictable influence, such as a reference to a song that is currently very popular.
To remove such outliers, for each label group, we compute the {\em z-score} of each post based on the influence score and remove those whose z-scores are larger than 2.

\subsection{Experiment design}

We first show the difference caused by using different grouping methods to separate the inherent meaning from the decorating parts of the post and choose a specific group-labeling method for model performance analysis.
Then we list the performance of the proposed model, the n-gram baseline model, and the tweet embedding model given the commercial posts.
In order to show more details on the attributes that affect the influence of the commercial tweets,
we conduct a feature-importance analysis on the proposed decoration model.
Finally, we set up a case study with a set of real commercial posts and apply the prediction model to the cases.

\subsection{Group label analysis}
\begin{table*}[htb]
\centering
\caption{simGroup\_emb group samples}
\begin{tabular}{c|c|c|l}
\hline
\textbf{\#}&\textbf{Size}&\textbf{Comment}&\multicolumn{1}{c}{\textbf{Example}}\\
\hline
\hline
\multirow{2}*{1}&\multirow{2}*{10360}&\multirow{2}*{Food-related}&Two simple ingredients, two very unique\\
&&&drinks. http://sbux.co/1nobwF9 \\
\hline
\multirow{3}*{2}&\multirow{3}*{22579}&\multirow{3}*{Stories \& Suggestions}&This pup doesn't have time to chase his\\
&&&tail, because he's too busy traveling\\
&&&the world http://yhoo.it/1WutqAR\\
\hline
\multirow{4}*{3}&\multirow{4}*{12651}&\multirow{4}*{Entertainment \& Events}&Happy birthday Emma Stone! We can't\\
&&&wait to celebrate with you at the\\
&&&\#AFIFest premiere of \#LALALAND 11/15.\\
&&&���� http://soc.att.com/LALALAND \\
\hline
\multirow{3}*{4}&\multirow{3}*{16062}&\multirow{3}*{Electronics \& Tech}&Learn how to securely mobilize your biz;\\
&&& join us at BlackBerry Mobility Summit\\
&&&Benelux http://blck.by/1LJDo2b \\
\hline
5&357&Special cases&\#blizzard2016 \#Jeep\\
\hline
\end{tabular}
\end{table*}
Because the model is built to predict the performance of a commercial post, the core meaning of the post is considered as fixed. 
Therefore we group the posts based on their core parts as described before, so that the prediction model can focus on the decoration parts of the tweet.
In the experiment, we first extract the key words for each tweet, and then apply three group methods to the posts: 
\begin{itemize}
\item {\em simGroup\_binary} featurizes the key words using a binary representation and cluster them with k-means++, 
\item {\em simGroup\_emb} featurizes the key words using Word2Vec provided by Gensim with pretrained word vectors \cite{rehurek_lrec}, 
then averages the word vectors to generate the vector for this tweet, and clusters them using kmeans++,
\item {\em topicGroup} applies Latent Dirichlet Allocation model \cite{blei2003latent} to get the topic distribution for each tweet, and groups the tweets based on the topic with highest probability.
\end{itemize}
After generating the groups of tweets, labels are assigned to each group individually and mixed together for the following prediction task.
Because the data size is limited, we test these different group-labeling methods with three, five, and seven groups.
In order to explore the difference caused by various labeling groups, we apply each labeling result to the n-gram model, proposed decoration model, and tweet embedding model as mentioned in the explanation of ``Separation from Inherent Meanings''.

Figure 2 shows the F1 score of three grouping methods with different numbers of groups given each prediction method.
Overall, {\em simGroup\_binary} and {\em simGroup\_emb} generate comparative performance, while {\em topicGroup} does not fit well for this task.
Furthermore, {\em simGroup\_emb} is more stable than the other two grouping methods with different numbers of groups.
As shown in the performances of decoration and embedding model, it is more suitable to use the pretrained word embeddings than the simple binary representation to group tweets.
The small portion of isolated key words included in the grouping process and the limited data size also justify the use of word embedding for tweet clustering.
Therefore, to get the best performance, we choose {\em simGroup\_emb} with five groups as the meaning separation method, and use the labels generated from it for the following experiment.
The performance of different prediction models will be further studied in the following section.

Table \RN{3} shows samples from the group assignment generated by {\em simGroup\_emb} with 5 groups.
Group 5 is a particular case where the posts only contain special elements such as hashtags and urls.
Unlike the other groups, the actual meaning of the hashtags may not fall into the same category. 
However, the limited size of this group ensures that this labeling process is still convincible.
On the other hand, the other four groups work as expected, so that the labeling process focuses on the decoration parts of the posts.
For example, the labeling within Group 1 represents how the construction and decoration of the posts is related to the influence of posts that carry food-related information.

We have also tried to group posts by other attributes such as its author account.

Accounts that belong to the same category are grouped together. For example, {\em Amazon, Google, and Yahoo} are grouped together as technology companies.
However, results show that this intuitive grouping method performs much worse than learning the clustering directly from the data.
In most cases, a single official account does not post commercials about only one type of products or promotions.
In fact, posting commercials through social media is far more flexible and easier than other approaches.
Therefore, companies tend to post a more comprehensive set of commercials through social platforms than traditional ones.

\subsection{Model performance and analysis}

After generating the labels using {\em simGroup\_emb} with five groups as described in the previous section, we use the full dataset for the performance test.
As described in Section 3.2.2 , we apply our proposed feature model to an SVM classifier with Radial Basis Function kernel.
We also apply the baseline approach with a MaxEnt Classifier, and the state-of-the-art tweet embedding model with the same SVM classifier for comparison.
To analyze the importance of the features, we conduct an ablation analysis on the proposed decoration features.
We are designing the system to help companies identify (or even craft) commercial tweets that are likely to have a large influence.
For this reason, we report the precision, recall, and F1 score for the positive cases.

\begin{table}[h]
\centering
\caption{Model performance and ablation analysis}
\begin{tabular}{l|c|c|c}
\hline
\textbf{Feature}&\textbf{Precision}&\textbf{Recall}&\textbf{F1}\\
\hline
\hline
Baseline(n-gram)&0.7597&0.7733&0.7664\\
\hline
Embedding&0.7616&0.8158&0.7878\\
\hline
Decoration(full)&0.7268&0.8708&0.7923\\
\hline
- Author meta&-0.0839&-0.1062&-0.0938\\
- Elements&-0.0097&-0.0032&-0.0071\\
- Punctuation&-0.0010&-0.0062&-0.0032\\
- Mentions&+0.0137&-0.0244&-0.0024\\
- Contain digit&-0.0013&-0.0027&-0.0019\\
- POS dist&-0.0005&-0.0006&-0.0006\\
- Sentiment&-0.0002&-0.0005&-0.0003\\
- Post meta&+0.0010&+0.0014&+0.0012\\
- Complexity&+0.0111&-0.0113&+0.0017\\
\hline
\end{tabular}
\end{table}

\begin{table*}[htb]
\centering
\caption{Prediction samples from different models where the true labels are positive}
\begin{tabular}{c|c|c|c|l}
\hline
&\textbf{N-gram}&\textbf{Embedding}&\textbf{Decoration}&\multicolumn{1}{c}{\textbf{Tweet}}\\
\hline
\hline
\multirow{3}*{1}&\multirow{3}*{1}&\multirow{3}*{1}&\multirow{3}*{0}&Community. Connection. Celebration. Today, and\\
&&&&every day. \#LGBTHistoryMonth \\
&&&&http://soc.att.com/2dAG6sI\\
\hline
\multirow{2}*{2}&\multirow{2}*{1}&\multirow{2}*{1}&\multirow{2}*{0}&Any terrain. Any season. Anytime.\\
&&&&pic.twitter.com/RnhHJWlvgF\\
\hline
\multirow{2}*{3}&\multirow{2}*{1}&\multirow{2}*{1}&\multirow{2}*{0}&Pro tip: Sweet bedding = sweet dreams. ��\\
&&&&http://mcys.co/2cx8pGf\\
\hline
\multirow{3}*{4}&\multirow{3}*{1}&\multirow{3}*{0}&\multirow{3}*{0}&Avocados + salt + lime + \_\_\_\_\_\_. What goes in your\\ 
&&&&guacamole? Super Fast Guac: http://bit.ly/1Y9oJOX \\
&&&&\#CincodeMayo\\
\hline
\multirow{3}*{5}&\multirow{3}*{0}&\multirow{3}*{1}&\multirow{3}*{1}&Due to forecasted winter weather in the Pacific\\
&&&&Northwest, we've issued a travel waiver for February\\
&&&& 3rd. More info: http://bit.ly/2iPzTuS\\
\hline
\multirow{3}*{6}&\multirow{3}*{1}&\multirow{3}*{0}&\multirow{3}*{1}&OBAP’s dedication to aspiring pilots inspires us, which\\
&&&& is why we're proud to support their programs that mold\\
&&&&the future of aviation.\\
\hline
7&0&1&1&Oh hey @trollhunters @Stranger\_Things\\
\hline
\multirow{2}*{8}&\multirow{2}*{1}&\multirow{2}*{0}&\multirow{2}*{1}&\#18thcenturyproblems \#PrideAndPrejudice\\
&&&&\#NowOnNetflix\\
\hline
\end{tabular}
\end{table*}

Table \RN{4} shows the performance of the baseline method, embedding model, and our proposed decoration model, as well as the variation in contribution of
the proposed features.
In general, the decoration model outperforms both n-gram baseline and the embedding model in terms of F1 score.
More specifically, the proposed model tends to have a much higher recall than the other two models, while a lower precision than the others.
The proposed decoration model does not include any particular meaning of the commercial post or the identity of the mentioned usernames and author account.
But the result shows that it has more capability to predict the potential influence of a commercial post than traditional content models such as n-gram and embedding models.
Moreover, without looking into the actual core content and the identities, it also reduces the risk of over-fitting the model to a specific dataset.
In this case, a more general model would work better on an unknown commercial post.

Further, our proposed feature set is much more compact than the n-gram features, and it is also more compact than the embedding model.
The decoration model is not only more general and adaptive,
but also more efficient and effective than the content-based n-gram model or embedding model in predicting successful commercial tweets.

Table \RN{5} lists the predicted labels from three models and the text of several sample tweets where the true label is positive (label 1).
More specifically, it shows the sample tweets where the decoration model has different predictions from the content models (n-gram or embedding model).

Most of the posts where the n-gram and embedding models correctly predict as positive while the proposed model does not are constructed in an informal way.
Many of them are not constructed as a complete sentence.
They are either the combination of several isolated words and phrases such as Tweet 1 and 2,
or written in a special form such as Tweet 3 and 4.
Although they have been adapted to the task of tweets, both the POS tagger and dependency parser are not able to work well on such incomplete sentences, which further affects the performance of the decoration model.
A bag-of-words assumption does not carry any order or dependency information; therefore it is less sensitive to these special cases.
Thus, the proposed decoration model generates a lower precision than the n-gram and embedding models.

On the other hand, the proposed model is able to successfully predict more positive cases than the other models.
We note that most tweets the proposed model predicts as positive while the n-gram or embedding models fail to predict as positive occur in two situations:
\begin{itemize}
\item The construction of the post is complicated, which usually means a complex sentence such as Tweet 5 and 6.
\item The major body of the post is built of special elements such as hashtags, urls or username mentions such as Tweet 7 and 8.
\end{itemize}
The complexity features and the structure analysis, such as tweet parsing, in the proposed model help locate and extract the posts that have positive influence.
Content-based n-gram and embedding models do not work well on longer and more complex sentences.

\begin{table*}[htb]
\centering
\caption{Commercial tweets about a raffle event for winning console controllers}
\begin{tabular}{c|l|c}
\hline
&\multicolumn{1}{c}{\textbf{Tweet}}&\multicolumn{1}{|c}{\textbf{Label}}\\
\hline
\hline
\multirow{3}*{1}&exclusive swag! starting tomorrow, you're entered to win a custom&\multirow{3}*{1}\\
&gecko-themed console controller every time you post using&\\
&\#GEICOGaming.&\\
\hline
\multirow{3}*{2}&love the \#GEICOGaming turnout. remember, every single post&\multirow{3}*{1}\\
&this weekend enters you to win an exclusive gecko-themed&\\
&console controller!&\\
\hline
\multirow{2}*{3}&every post (!) using \#GEICOGaming this weekend makes you&\multirow{2}*{0}\\
&eligible for a custom console controller. bring it!&\\
\hline
\multirow{3}*{4}&starting tonight at midnight, every social post containing&\multirow{3}*{0}\\
&\#GEICOGaming enters you to win a custom console controller!&\\
&get in while you can.&\\
\hline
\hline
\multirow{3}*{5}&exclusive swag, limited opportunity! every post (!) using&\multirow{3}*{1}\\
&\#GEICOGaming this weekend makes you eligible for a custom&\\
&console controller. bring it!&\\
\hline
\multirow{3}*{6}&check this great opportunity! starting this midnight, every social&\multirow{3}*{1}\\
&post containing \#GEICOGaming enters you to win a custom&\\
&console controller!&\\
\hline
\end{tabular}
\end{table*}

As expected, the ablation analysis shows that the author meta feature has the biggest impact on the final prediction in terms of F1 score.
The special elements used in the post are other attributes that contribute meaningfully to the final prediction.
They are very common and useful in commercial tweets.
Moreover, the mentioned usernames and types of punctuation have considerable impact as well.
The use of these two attributes is also more popular and effective in commercial posts than regular tweets.
Complexity is shown to have an impact in generating a higher recall; we have the same result from the previous analysis.
The sentiment feature is shown to be less of a differentiator. 
As mentioned before, commercial tweets are written to be non-negative, and most commercial tweets have very limited sentiment difference. 
Finally, we found the post meta feature working in an unexpectedly way, such that removing this feature improves the model.
This shows that the posting time of commercial tweets is not as useful as the posting time of regular tweets \cite{artzi2012predicting}.

\section{Demonstrating Use of the Framework: A Case Study}
To demonstrate a real use of our prediction model, we pick four commercial tweets posted by {\em GEICO} ($1$ through $4$),
and two modified tweets (5 and 6).
These tweets, about a raffle event in which one can win a console controller, are shown in Table \RN{6}.
The label column lists the prediction from the proposed model, and they all agree with the true labels for the real tweets ($1$ as positive and $0$ as negative).
We exclude the post meta feature to ensure the consistency across all cases.

The four real tweets deliver the same core information about the raffle.
However, they differ in their success in generating influence.
The positive cases include additional phrases before the core information that serve as hooks to raise readers' interest.
Our model is able to correctly capture that such hooks are indeed effective.

The positive real tweets are found to have higher readability scores than the negative ones,
mainly owing to the use of additional phrases and subtler use of words.
Although a higher readability score generally implies that the tweet is more difficult to follow,
it can also mean a more precise and attractive expression of the message.
The sample tweets show a positive impact of such an expression on the influence of the tweets.
In addition, we note that the positive cases contain a greater number of nouns than verbs.
Although nouns such as ``swag'' and ``turnout'' do not contain core information, they are useful in drawing more attention.

Samples Tweets $5$ and $6$ are created from samples Tweets $3$ and $4$, with the addition of certain hook phrases to the beginning of the posts.
Minor changes are also made to the main body to meet the length limitation.
These modifications lead to an increase in the number of parse tree heads,
and an increase in readability and sentiment scores as well.
With these modifications, the proposed system predicts that the modified tweets will have a positive influence.
In other words, these changes help the tweets have more influence while still conveying the same information.

The above case study shows a successful use of the system to predict the influence of real commercial posts.
Most pertinently, it shows that one can use the system to craft a tweet till it is predicted to be successful.
\section{Conclusions}
This paper describes a comprehensive feature model to predict the potential influence of a commercial tweet to its readers.
The proposed model does not include the inherent meaning of the post and it relies on only the construction, decoration and meta feature of the post.
It also ensures the generality of the model such that it can be adapted to most commercial posts.
Unlike some previous work, the model does not need any observation of the diffusion of the post,
and therefore the prediction can be made instantly before posting a tweet.
The experiments show that our techniques can provide a useful and stable performance in predicting the tweets with successful influence while using only a small set of features.
The proposed decoration model outperforms the content-based n-gram and embedding models in terms of F1 score.
We also show that among all the features, author meta data has the largest contribution, 
while the special elements, punctuation marks, and username mentions contained in the post have comparable contribution as well.

On the basis of this prediction system, we will also likely develop a suggestion system that can help companies generate better commercial tweets in terms of the influence on readers.
Similar to the example shown in the case study, the suggestion system can propose potential modifications, using the prediction system to determine which modifications lead to a more successful post.
Other techniques can be explored to separate the writing of the tweet from the commercial information it carries.
Working on a given informational basis is an essential task when advertising through social platforms.

\bibliographystyle{IEEEtran}
\bibliography{reference}

\end{document}